\documentclass[11pt,preprint]{aastex}

\usepackage{subfigure}
\usepackage{epsfig}
\def \arcsec {$^{\prime\prime}$}
\def \arcmin {$^{\prime}$}

\def \msols {M$_{\odot}$}

\shortauthors{Anderson et al.}

\begin{document}

\title{From Gas to Stars in Energetic Environments: Dense Gas Clumps in the 30 Doradus Region Within the Large Magellanic Cloud}

\author{Crystal N. Anderson$^{1}$, David S. Meier$^{1,2}$, J\"urgen Ott$^{2,1}$, Annie Hughes$^{3}$,Tony Wong$^{4}$, Christian Henkel$^{5,6}$, Rosie Chen$^{5}$, Remy Indebetouw$^{7,8}$,Leslie Looney$^{4}$, Erik Muller$^{9}$, Jorge L. Pineda$^{10}$, Jonathan Seale$^{11}$}

\affil{$^{1}$Physics Department, New Mexico Institute of Mining and Technology, 801 Leroy Pl., Socorro, NM 87801, USA}
\affil{$^{2}$National Radio Astronomy Observatory, P.O. Box O, Socorro, NM 87801, USA}
\affil{$^{3}$Max-Planck-Institut f\"ur Astronomie, K\"onigstuhl 17,  69117, Heidelberg, Germany}
\affil{$^{4}$Astronomy Department, University of Illinois, 1002 W. Green Street, Urbana, IL 61801, USA}
\affil{$^{5}$Max-Planck-Institut f\"ur Radioastronomie, Auf dem H\"ugel 69,
D-53121 Bonn, Germany}
\affil{$^{6}$ Astronomy Department, King Abdulaziz University, P.O. Box 80203, Jeddah 21589, Saudi Arabia}
\affil{$^{7}$National Radio Astronomy Observatory, 520 Edgemont Rd, Charlottesville, VA 22903, USA}
\affil{$^{8}$Department of Astronomy, University of Virginia, P.O. Box 3818, Charlottesville, VA 22903-0818, USA}
\affil{$^{9}$National Astronomical Observatory of Japan, 2-21-1 Osawa, Mitaka, Tokyo 
181-8588, Japan}
\affil{$^{10}$Jet Propulsion Laboratory, 4800 Oak Grove Dr,  Pasadena, CA 91109, USA}
\affil{$^{11}$Space Telescope Science Institute, 3700 San Martin Drive,Baltimore, MD 21218, USA}

\begin{abstract}
We present parsec scale interferometric maps of HCN(1-0) and HCO$^{+}$(1-0) emission from dense gas in the star-forming region 30 Doradus, obtained using the Australia Telescope Compact Array. This extreme star-forming region, located in the Large Magellanic Cloud (LMC), is characterized by a very intense ultraviolet ionizing radiation field and sub-solar metallicity, both of which are expected to impact molecular cloud structure. We detect 13 bright, dense clumps within the 30 Doradus-10 giant molecular cloud. Some of the clumps are aligned along a filamentary structure with a characteristic spacing that is consistent with formation via the varicose fluid instability.  Our analysis shows that the filament is gravitationally unstable and collapsing to form stars. There is a good correlation between HCO$^{+}$ emission in the filament and signatures of recent star formation activity including H$_{2}$O masers and young stellar objects (YSOs). YSOs seem to continue along the same direction of the filament toward the massive compact star cluster R136 in the southwest. We present detailed comparisons of clump properities (masses, linewidths, sizes) in 30Dor-10 to those in other star forming regions of the LMC (N159, N113, N105, N44). Our analysis shows that the 30Dor-10 clumps have similar mass but wider linewidths and similar HCN/HCO$^{+}$ (1-0) line ratios as clumps detected in other LMC star-forming regions. Our results suggest that the dense molecular gas clumps in the interior of 30Dor-10 are well-shielded against the intense ionizing field that is present in the 30Doradus region.  

\end{abstract}


\section{Introduction}

Throughout the years many studies on Galactic star formation have been conducted (e.g. Kennicutt \& Evans 2009; McKee \& Ostriker 2007; Zinnecker \& Yorke 2007). It has been observed and theorized that within giant molecular clouds (GMCs) there are clumps and small cores. This resulted in the idea that GMCs are hierarchical in nature with substructures spanning a large range of sizes. Clumps are dense regions with masses $\sim$10$^{2}$-10$^{3}$M$_{\odot}$ and sizes of a few parsec, whereas cores are smaller ($\sim$0.1\,pc) and denser. It is here where individual star formation occurs (Williams et al. 2000).

The physical processes that determine how molecular clouds fragment, form clumps/cores and then stars depends strongly on the past history of star formation. This occurs through both recent radiative and mechanical feedback from massive stars and, on longer term, through enhanced cooling due to the buildup of metals. Radiative and mechanical energy input from stellar populations can alter subsequent star formation over a large part of a galaxy and is hence relevant to the evolution of galaxies. Much of our knowledge of star formation on galaxy wide scales is based on scaling laws and other parametric descriptions of star formation (e.g. Kennicutt 2012). But to understand the overall evolution of star formation in galaxies we need to watch the feedback processes at work on giant molecular cloud (GMC) scales. By doing this we can begin to answer how strong feedback environments change the properties of the substructure in GMCs.

Applying the theory of Galactic star formation to other galaxies has been a challenging process due to the lack of resolution with current instruments. However, only the nearest galaxies allow us to resolve GMCs and their substructures. The Large Magellanic Cloud (LMC), is one of the closest galaxy (D$\sim$ 50\,kpc) and is close enough that current instruments can resolve the sub-structure of its GMCs to $\lesssim$1pc.

Molecular gas is seen across the disk of the LMC, with previous studies revealing a dominant ridge of bright CO emission extending $\sim$ 2\,kpc south of 30 Doradus, the most extreme star forming region in the Local Group (e.g. Columbia 1.2\,m Millimeter Wave Telescope, Cohen et al. 1988; NANTEN, Fukui et al. 1999; MAGMA survey with the Australia Telescope National Facility [ATNF] Mopra telescope, Ott et al. 2008, Pineda et al. 2009, Muller et al. 2010, Hughes et al. 2010, Wong et al. 2011). Despite its proximity ($\sim$ 20\,pc) to the young star cluster powering 30 Doradus, R136, and its intense radiation field ($\chi \sim$500$\chi_{0}$ in units of Draine [1978]; Pineda et al. 2009), a molecular cloud, 30Dor-10 is observed towards the HII region (e.g. Johansson et al. 1998).  CO (1-0) observations of this GMC reveal that it is about 40\,pc in size, fainter, more compact and has broader linewidths than other LMC clouds (Johansson et al. 1998; Heikkil\"a et al. 1999; Rubio et al. 2009).  Recent Atacama Large Millimeter Array (ALMA) observations of 30Dor-10 in $^{12}$CO, $^{13}$CO,  and C$^{18}$O probe the highest resolution view of molecular gas in this extragalactic star forming region ($\sim$\,0.4\,pc $\times$ 0.6\,pc) (Indebetouw et al. 2013). The $^{12}$CO is mostly associated with clumps and small 'pillars of creation'. This makes 30Dor-10 a perfect target to study extragalactic GMC dense gas properties toward a sub-solar metallicity environment under the influence of an intense starburst.

Other active star forming GMCs have been observed at lower resolution along with 30Dor-10 using SEST, such as the N159 HII complex (Johansson et al. 1994; Chin et al. 1997) in the LMC. These observations report the detection of eight molecules ($^{12}$CO, CS, SO, CN, HCN, HNC, HCO$^{+}$, and H$_{2}$CO). Heikkil\"a et al. (1999) detected nine molecules ($^{13}$CO, $^{12}$CO, CN, CS, HNC, HNC, HCO$^{+}$, C$_{2}$H, and C$_{3}$H$_{2}$) and Wang et al. (2009) detected several molecules toward the star forming regions N113, N44, and N214. Complementary high resolution interferometric observations of a subset of these GMCs were carried out in the dense gas tracers HCN(1-0), HCO$^{+}$(1-0), and NH$_{3}$ (1,1) and (2,2) towards selected non-30Doradus GMCs in the LMC using the Australia Telescope Compact Array (ATCA) (Wong et al. 2006; Ott et al. 2010; Seale et al. 2012, hereafter S12).

In this paper, we extend the S12 sample (e.g. N159, N113, N105 and N44) to include 30Dor-10, to probe dense gas that is embedded in the most extreme radiation field in the LMC.  In section 2, we describe our observational strategy and data reduction. In section 3, we analyze the physical properties of the HCN and HCO$^{+}$ clumps in 30Dor-10, and compare them to the properties of HCN and HCO$^{+}$ clumps detected in other LMC GMCs by S12. In section 4, we discuss a potential origin for the filamentary morphology of the dense gas tracers in 30Dor-10, and compare our maps to the higher resolution ALMA data recently published by Indebetouw et al. (2013). Our key results and conclusion are summarized in section 5.

\section{ Observations and Data Reduction}

 We observed the dense gas cloud, 30Dor-10 with the ATCA\footnote{The Australia Telescope Compact Array is part of the Australia Telescope National Facility which is funded by the Commonwealth of Australia for operation as a National Facility managed by CSIRO.}.  HCO$^{+}$ (J=1$\rightarrow$0; 89.1885\,GHz rest frequency) and HCN(J=1$\rightarrow$0; 88.6318\,GHz rest frequency)  are used to probe the structure of the GMC. Ten separate observations were taken during 2011 September 16-28. We observed 28 pointing centers, which were mosaicked together to cover the inner 3\arcmin \,of the 30Dor-10 region. The Compact Array Broadband Backend (CABB) correlator was used to observe the redshifted frequencies of HCO$^{+}$ and HCN simultaneously, so observations from both transitions have consistent sensitivities and ($u,v$) coverage.  A bandwidth of 2\,GHz with 2048 x 1\,MHz channels and up to 16 zoom bands in each IF band were used for the CABB correlator setup. For a given zoomband, the native spectral resolution of the correlator configuration corresponds to a channel width of 0.5\,kHz (1.6\,m $\rm s^{-1}$), but we binned the channels to a velocity width of 0.84\,km \rm s$^{-1}$ to increase the signal to noise prior to analysis. Typical clump linewidths are $\sim$ 3\,km s$^{-1}$ (see Table 1), so the clumps remain spectrally well-resolved at this spectral resolution. The center of the band was tuned to a redshifted velocity of 255\,km s$^{-1}$ (LSRK), which is the average velocity of the area. All observations were conducted using the most compact array configuration (H75) with baselines of 31-48\,m giving a synthesized beam FWHM of 6.8\arcsec $\times$ 5.8\arcsec (1.6\,pc $\times$ 1.4\,pc at the assumed distance of $\sim$ 50\,kpc to the LMC).  Complex gain calibration was performed using PKS 0637-752 for all observations and PKS 2223-052 was used to calibrate the bandpass.  Pointing was measured every hour on the phase calibrator with an accuracy of 2-5\arcsec.  Flux calibration was performed using Uranus. The absolute flux calibration uncertainty is estimated to be $\sim$50\%. However, relative calibration between spectral lines should be significantly better, $\lesssim$10\%. Data were reduced, imaged and cubes created using ATNF MIRIAD software package (Sault et al. 1995).  The root mean square (rms) of the brightness fluctuations due to noise for both HCO$^{+}$ and HCN cubes is 10\,mJy/beam in a 0.84\,km s$^{-1}$ channel. 

\section{Results}

We examine the internal structure of the extragalactic GMC in the star forming cloud 30Dor-10 at 1.45\,pc resolution. Figures 1a and 1b show maps of the integrated intensity of HCO$^{+}$ (1-0) and HCN(1-0) data while Figure 1c shows the peak brightness temperature map of HCO$^{+}$ (1-0) with clump regions in ellipses overlaid, clump region detection is discussed in more detail in the section below. Figure 2 displays an image of 30Dor-10 indicating where intense star formation is occurring.  H$\alpha$ is in green, 8$\micron$ in red, and X-ray emission in blue. The dense gas morphology of 30Dor-10 together with H$_{2}$O masers and YSO counterparts (Gruendl et al. 2009) are overlaid on Figure 2.

We observe a roughly linear structure extending outward from the star cluster R136 in the south-west, which bifurcates near the center of the field.  This structure is resolved into a collection of dense clumps. We detect 13 clumps in HCO$^{+}$(1-0), the four brightest of which are also detected in HCN(1-0) and are associated with previously identified YSOs.  Properties of individual clumps are determined using CPROPS (see section 3.1) and can be found in Table 1.  H$_{2}$O masers are also coincident with the brightest HCO$^{+}$ peak. Determined either from virial masses or clumps' masses under local thermal equilibrium (LTE), the masses span from 10$^{2}$ to 10$^{3}$\,M$_{\odot}$. CPROPS resolves 4 of the 13 clumps with sizes of 0.9 - 1.6 pc, the remaining clumps are unresolved with radii $<$0.7\,pc. These sizes are consistent with the other GMCs in the LMC discussed in S12. The GMCs are significantly brighter in HCO$^{+}$(1-0) than HCN(1-0), with typical line ratios HCN/HCO$^{+}$$\sim$ 0.20 for the GMCs with M$>$10$^{3}$M$_{\odot}$.

\subsection{Clump Identification}

We use HCO$^{+}$  as the primary clump tracer, since HCN is much weaker. To identify HCO$^{+}$ clumps in the ATCA data we used the CPROPS procedure presented by Rosolowsky \& Leroy (2006) (R06).  We follow the method of Wong et al. (2011) hereafter W11 and S12, respectively to facilitate direct comparison with these studies. We briefly review the basic properties of the algorithm here. 

CPROPS uses a dilated mask technique to isolate clump regions of significant emission within spectral line cubes.  Clump regions of significant emission in the ATCA data are identified by finding volume pixels ("voxels") with emission greater than a 3$\sigma$ threshold. CPROPS is run on position-position-velocity data cubes where the cloud is continuous and bordered by an isosurface of brightness temperature, T$_{\rm edge}$. Clump regions are determined when the pixel temperature in the cloud is $>$T$_{\rm edge}$ and that the pixels outside the cloud have T$<$T$_{\rm edge}$.

Once the clump regions are determined, we apply various criteria to estimate radii, line widths and other parameters, as follows. We require that each region has an area larger than two beam sizes and a velocity width of more than a single channel. Also, the intensity contrast between the peak and the edge of the region must be greater than a factor of two. If it meets the above criteria then the cloud can be decomposed, otherwise it is considered noise. Local maxima are then determined by identifying the elements in the data cube that are larger than their neighbors. Once the local maxima are found then these are labeled as a clump and the parameters are then extracted for each clump as described in R06. 

The quoted radius is determined by the geometric mean of the second spatial moments along the minor and major axis. Also, any flux detected below the threshold is assumed to be properly modeled by the extrapolation algorithm. The clump radius is determined by assuming a brightness profile and distance. If the extrapolated clump size is smaller than the synthesized beam, it cannot be deconvolved. Marginally resolved clumps suffer from substantial radii uncertainties (S12). 

\subsubsection{Clump Parameters}
Following, S12, we derive the same clump properties i.e linewidth ($\Delta v$), virial mass (M$_{\rm vir}$), LTE mass (M$_{\rm LTE}$), and radius. We compare our values to what S12 obtained for the other GMCs in the LMC in Table 2. The velocity dispersions are found in the same way as the radii are, except they are extrapolated to what would be measured in zero-noise observations and then deconvolved from the channel width.  Assuming Gaussian line profiles, the linewidth is determined from the full-width at half maximum as in S12. The flux is determined in a similar manner but by using the zeroth moment. The HCO$^{+}$ luminosity is determined from the integrated flux scaled by the square of the distance (W11):

\begin{equation}
L_{\rm mol}[\rm K\,km\,s^{-1} pc^{2}] = D^{2} \displaystyle \sum_{\rm i} T_{\rm i}\delta V,
\end{equation}

 \noindent where $\delta \rm V$ = $\delta x \times \delta y \times \delta \rm v$ is the volume of the voxel, T$_{\rm i}$ denotes brightness temperatures of individual voxels within the region, D is the distance to the LMC in parsec.

The virial mass is determined by:

\begin{equation}
M_{\rm vir}[M_{\odot}]=1040\sigma_{\rm v}^{2}\rm R
\end{equation}

\noindent where $\sigma_{\rm v}$ is the velocity dispersion in km\,s$^{-1}$ and R is the radius in pc. This assumes a cloud with a radial density profile of $n \propto r^{-1}$. Virial masses are shown in Table 1 along with the H$_{\rm 2}$ gas mass traced by HCO$^{+}$ under LTE  conditions, which is determined by following equation (9) from Barnes (2011):

\begin{equation}
M_{\rm LTE}(\rm H_{2})=\frac{m_{\rm H_{2}}\pi R^{2}}{X_{\rm HCO^{+}}}N_{\rm HCO^{+}}
\end{equation}

\noindent where N$_{\rm HCO^{+}}$ is the column density determined by assuming optically thin emission and LTE, shown in equation (4), X$_{\rm HCO^{+}}$ is the HCO$^{+}$ abundance relative to H$_{2}$, m$_{\rm H_{2}}$ is the mass of molecular hydrogen, and R is the radius of the individual clumps in pc. 

\begin{equation}
N_{\rm HCO^{+}}(\rm cm^{-2}) = 5.79\times 10^{10}~ ~T_{\rm ex} e^{4.28/T_{\rm ex}} I_{\rm HCO^{+}}(\rm K\,km\,s^{-1}),
 \end{equation}

\noindent where T$_{ex}$ is the excitation temperature. The HCO$^{+}$ column density is derived with molecular parameters from the Splatalogue\footnote{http://splatalogue.net} database. An excitation temperature (T$_{\rm ex}$) of 30\,K is adopted which is typical for massive clumps observed in dust continuum (e.g. Fa\'undez et al. 2004; Beltr\'an et al. 2006) and a HCO$^{+}$ abundance relative to H$_{2}$ of X$_{\rm HCO^{+}} \sim 10^{-9}$ is assumed (Barnes et al. 2011). If the HCO$^{+}$ abundance relative to H$_{2}$ is increased, the H$_{2}$ gas mass traced by HCO$^{+}$ will decrease by a corresponding factor. On the other hand, an increase in the excitation temperature to T$_{\rm ex} \sim$ 50\,K will increase the mass by a factor of two. All clump parameters discussed above are found in Table 1.

\subsection{Clumps Properties}

ATCA observations similar to ours were performed on the star forming regions N159, N105A, N44, and N113 by S12, who detected individual clumps with the same dense gas tracers HCO$^{+}$(1-0) and HCN(1-0). S12 used the same compact antenna array configuration of the ATCA, CABB correlator settings which result in a very similar spatial and spectral resolution and sensitivity compared to our observations.  Compared to these GMCs, 30Doradus exhibits a much higher star formation rate (SFR), more intense and extended UV flux, and is forming a super-star cluster. N159 is located south of 30Dor-10 along the molecular ridge (Pineda et al. 2009, 2012) and has a high local radiation field due to individual star forming regions. N113, N105A and N44 have progressively weaker interstellar radiation field, respectively. The individual GMC properties are recorded in Table 2.

In the following sub-sections we describe both the clump properties of 30 Dor-10 and how they compare to the other LMC GMCs. We investigate dense-gas fraction, peak temperature-linewidth, RMS centroid velocity-CO linewidth, total CO mass to total HCO$^{+}$ mass, and HCN/HCO$^{+}$ line ratio relations as shown in Figure 3. Since CPROPS only obtained deconvolved radii for four of the 13 clumps, the size-linewidth and virial mass-linewidth could not be reliably determined for the entire sample.

\subsubsection{Fraction of Dense Gas}

We compare the H$_{2}$ mass traced by HCO$^{+}$ of all the clumps in each star forming region (N159, N113, N105A and N44) to the molecular gas masses estimated from the GMCs' CO luminosities from MAGMA data. The MAGMA CO data (Figure 2) covers our entire observed region unlike the ALMA map and we therefore use MAGMA for calculations of the bulk properties of the entire GMC. M$_{\rm CO}$ is calculated from W11 equation (2):
 
\begin{equation}
M(\rm H_{2})_{CO}[M_{\odot}]=4.4\frac{X_{\rm CO}}{2.2\times 10^{20}cm^{-2}(\rm K\,km\,s^{-1})} L_{\rm CO},
\end{equation}

\noindent where the Galactic CO-to-H$_{2}$ conversion factor

\begin{equation}
 X_{\rm CO}\simeq 2.0\times 10^{20} \rm cm^{-2}(\rm K\,km\,s^{-1})^{-1},
 \end{equation} 
 
\noindent is adopted (Strong et al. 1988) but see Bolatto et al. (2013) for other conversion factors. $L_{\rm CO}$ is the CO luminosity and values are shown in Table 2 for each cloud.  $L_{\rm CO}$ comes from the MAGMA CO data which is measured directly from the integrated intensity map. The CO allows us to examine how total molecular gas compares to the dense gas in each individual GMC.  The dense gas fraction for 30Dor-10 (as measured by HCO$^{+}$(1-0)/CO(1-0)) is quite similar to the other GMCs in the LMC. 30Dor-10's dense gas fraction is $\sim$1/12 whereas the other mapped GMCs in the LMC have a dense gas fraction $\sim$ 1/10.  

\subsubsection{M$_{\rm vir}$ versus M$_{\rm LTE}$}

Each clump's virial mass and LTE molecular mass, traced by HCO$^{+}$, is found in Table 1.  If the LMC clumps are composed entirely of dense gas and are in gravitational equilibrium then we would expect M$_{\rm vir}$/M$_{\rm LTE}$ $\sim$ 1. The spatially resolved clumps in 30Dor-10, however, exhibit a ratio of  M$_{\rm vir}$/$\rm M_{\rm LTE}$ $\sim$6. Such a high ratio may be attributed to  several possibilities, including 1) the adopted T$_{\rm ex}$ is underestimated, 2) the HCO$^{+}$ abundance of X$_{\rm HCO^{+}} \sim$ 10$^{-9}$ could be overestimated, 3) the dense gas traced by HCO$^{+}$ accounts for less of the total clump mass,  or 4) that the virial mass estimate is incorrect and clumps are slightly super-virial.  On the other hand, compared to 30Dor-10 the other GMCs also have elevated $\sum$M$_{\rm vir}$/$\sum \rm M_{\rm LTE}$, but by a smaller amount (a factor of three).  

The LTE mass determined is dependent on the chosen T$_{\rm ex}$ and HCO$^{+}$ abundance and so is likely the largest source of uncertainty.  Therefore we consider options 1) and 2) the most likely cause of the overall difference between M$_{\rm vir}$ and M$_{\rm LTE}$ in the LMC.  However the larger discrepancy between M$_{\rm vir}$ and M$_{\rm LTE}$ in 30 Dor-10 suggests that some extra reason is needed to explain the factor of three difference.  Given 30 Dor-10's more intense UV radiation field and dynamical impact of star formation, both higher gas excitation and clumps being super-virial likely are the causes of the increase (see section 3.2.3).

\subsubsection{Velocity Distribution}

Figure 3c, displays the RMS of the velocity centroids of the individual dense clumps versus the CO linewidth of the parent GMCs and provides an indication of how the dense clumps are moving relative to the more diffuse molecular material of the whole GMC. Within the errors the dense clumps in 30 Dor-10 fully sample the GMC's CO velocity distribution and so the dense gas is not dynamically segregated from the less dense gas.  

For a given brightness, the clumps in 30Dor-10 have a larger line width relative to the clumps in the other four GMCs (Figure 4).  The clumps in 30Dor-10 lie at the bottom of the distribution having a weaker temperature and broader linewidths. The clump line widths of the rest of the four GMCs exhibit similar behavior with respect to each other.  A weak trend for increasing clump brightness with line width in 30 Dor-10 is also observed.  Evidently 30Dor-10 is significantly more turbulent than the other GMCs.

\subsubsection{L(HCN)/L(HCO$^{+}$) Line Ratio}
  
The L(HCN)/L(HCO$^{+}$) line ratio relates the two commonly used dense gas tracers. Figure (3d) shows the line luminosity ratio determined from CPROPS for comparison between these five GMCs in the LMC. The GMCs are significantly brighter in HCO$^{+}$(1-0) than HCN(1-0), with typical line ratios L(HCN)/L(HCO$^{+}$)$\sim$ 0.20 for the GMCs with M$>$10$^{3}$M$_{\odot}$. 30Dor-10 has a similar line ratio as the other GMCs. 

\section{Discussion}

\subsection{30Dor-10 ALMA and ATCA Comparison}

We compare our ATCA HCO$^{+}$ dense gas clumps to the ALMA CO data,  to investigate the physical properties of 30Dor-10. Compared to the ALMA clump decomposition, we detect 7 HCO$^{+}$ clumps over the mutually observed field of view (FOV) while the ALMA data resolve 103 CO sub-clumps. Comparing the morphology of CO to that of HCO$^{+}$ we observe that the two are very similar in structure.  Over the ALMA field, HCO$^{+}$ is detected from all prominent CO clumps and the brightest CO clumps tend to be the brightest CO clumps.  The only exception is clump 10, which is brighter in HCO$^{+}$.  However the integrated intensity map of $^{12}$CO shows more diffuse, continuous emission than what we see in the HCO$^{+}$ data (Figure 5). 

The virial masses of the detected clumps in the ALMA and ATCA data are both
$\sim$10$^{3}$\msols, providing a good indication of the amount of mass
that is gravitationally bound in the associated clumps. On a clump by clump
basis the ratio of $^{12}$CO/HCO$^{+}$ is nearly constant for the filament
being around $\sim$ 12 - 14.5. Only clump 3, which is one of farthest from
R136 covered in the ALMA field, has a significantly higher ratio
($\sim$30). The $^{12}$CO/HCN ratios are much higher across the whole GMC
($\sim$60 - 140 for the filament, up to ~300 for clump 3), consistent
with the low HCN/HCO$^{+}$ ratio discussed in section 3.2.4.  Clearly
from these ratios the filament is the densest part of the cloud.  The
clump by clump ratios of 12 - 14.5 is very similar to what we see with
the $^{12}$CO/HCO$^{+}$ MAGMA data (section 3.2.1), giving further
evidence that the large scale measurements are being dominated by
clump/filament properties and not by a diffuse CO medium (see section
3.2.3).

\subsection{Filaments}

The HCO$^{+}$ velocity profile of the linear feature can be determined by taking a position-velocity slice along its structure (Figure 6).  The position-velocity diagram shows almost no change in velocity, hence the clumps are either moving uniformly along the structure with the same speed and direction or its in the plane of the sky.

The dense gas in 30Dor-10 may potentially be a portion of a filament extending outward from the star cluster, R136. A tight connection between filaments and star formation in Galactic dense cores has been revealed by the Herschel Gould Belt Survey (Andr\'e et al. 2010). This survey supports an emerging picture of star formation, in which thin, long ($>$\,pc scales) filaments form first a molecular cloud and then the densest part of the filaments fragment into protostellar cores via gravitational instability (Inutska et al. 1997). 

For non-star forming regions, like Polaris, all filaments have subcritical mass per unit length. This gives an explanation for the extinction threshold for the formation of prestellar cores of Av $\sim$ 7 observed in Aquila and Polaris.  Andr\'e et al. (2010) were able to determine that the extinction threshold corresponds to the threshold above which interstellar filaments are gravitationally unstable and can collapse to form stars.   They found that $>$ 60\% of the bound prestellar cores and Class 0 protostars in Aquila are concentrated in gravitationally unstable filaments where the mass per unit length exceeds the critical mass per unit length (eqn. 6) required for hydrostatic equilibrium.

While 30Dor-10 is significantly more distant than the molecular clouds studied by Andr\'e et al. (2010), the structure of the bulk of HCO$^{+}$ emission in 30Dor-10 bears a resemblance to a filament (Figures 1 and 2). This structure has a Y-shaped morphology, which could be one bifurcated filament or two filaments interacting. If the structure is a filament, we consider whether it is dense enough to be gravitationally unstable.  The clump critical mass per unit length depends  only on the temperature (Ostriker 1964),  

\begin{equation}
\left(\frac{M}{L}\right)_{\rm crit}=2c^{2}_{s}/G\sim 15\,T_{10}\,M_{\odot}\,\rm pc^{-1} 
\end{equation}
where $c_{s}$ is the isothermal sound speed, and $T_{10}$ is the gas temperature in units of 10\,K. 

We estimate the filament's observed clump mass per unit length. Adopting $T=30$\,K for dense clumps (section 3.1.1) we get a clump critical mass per unit length for the filamentary structure detected in 30Dor-10 of $\left(\frac{M}{L}\right)_{\rm crit} \sim 45~ M_{\odot}$/pc. From our observations and the CPROPS algorithm we know the length of the filamentary structure and the total calculated molecular gas mass for each clump. With these values the virial mass per unit length of M$_{vir}/L \sim 10^{3}~ M_{\odot}$/pc which is much larger than the critical clump mass per unit length.  Even adopting instead, M$_{\rm LTE}/L$ and T$\lesssim$100 K as extremes, the filamentary structure would still be above the critical mass per unit length.  This confirms the filamentary structure is gravitationally unstable and can collapse to form stars. 

The linear feature is indeed forming stars and there is some evidence that star formation appears to evolve along the structure.  As mentioned in Indebetouw et al. (2013), the compact star cluster K1 in the LMC is located at the tip of what they refer to as "pillars of creation" (clump 4, Figure 1(d)). In HCO$^{+}$(1-0), we observe the dense gas associated with this 'pillar' as a filament extending outward from the star cluster, R136, continues back into the CO cloud (Figure 2).  Toward 30Dor-10, H$_{2}$O masers and YSOs tends to correlate with the positions of HCO$^{+}$ clumps along the filament.   Additional YSOs are observed to extend back toward R136 along the same line. The presence of YSOs and H$_{2}$O masers demonstrates that HCO$^{+}$ cores are currently forming stars. This suggests that the 30Dor-10 clumps may be slightly younger than those near R136 and we are seeing the dispersal of molecular gas in that part of the filament. 

The clumps in Figure 1, do not appear to be randomly spaced. The clumps appear to be spaced sparsely but roughly uniformly along the filament. The idea that clumps along a filament are uniformly spaced is discussed in Jackson et al. (2010) for the Galactic Infrared Dark Cloud (IRDC) known as the "Nessie" nebula. These authors suggest stellar cluster formation arises from the fragmentation of filaments due to a varicose fluid instability.  In the Jackson et al. (2010) picture, the clumps should have spacings equal to the wavelength ($\lambda_{\rm max}$) of the fastest growing unstable mode of a self-gravitating fluid.  For the case of an infinite isothermal gas cylinder $\lambda$ corresponds to 22 H$_{e}$, where H$_{e}$ is the effective scale height,

\begin{equation}
H_{\rm e} = \frac{\Delta \rm v}{\left(4\pi G \rho_{c} \right )^{1/2}},
\end{equation}
and $\Delta \rm v$ is the linewidth. A self-gravitating fluid usually assumes thermal pressure to be the dominant source of gas pressure. However, in our case turbulent pressure dominates over thermal ($\Delta \rm v >>$c$_{s}$; Figure 3c). The average value of the linewidth for the clumps in 30Dor-10  is 3.72\,km s$^{-1}$, and assuming the same central volume density as in the "Nessie" filament, $\rho_{c}$$\sim$10$^{4}$\,cm$^{-3}$ we get a $\lambda_{\rm max}$$\sim$5.7\,pc. Observationally, the clumps along the filament are spaced $\sim$ 4.8-7.0\,pc apart which is consistent with the implied $\lambda_{\rm max}$.

In summary, the morphology and dynamics appear consistent with the string of clumps in 30Dor-10 being a gravitationally unstable filament with some evidence that the phase of star formation occurring in it evolves. 

\subsection{L(HCN)/L(HCO$^{+}$) and Dense Gas Properties}

Changes in the L(HCN)/L(HCO$^{+}$) intensity ratios from unity found in high metallicity starburst galaxies have been variously interpreted as due to changes in dense gas density (Meijerink et al. 2007; Baan et al. 2008; Krips et al. 2008; Meier \& Turner 2012) or to changes in the nature of the ionization rate or source (e.g. UV, X-rays, cosmic rays) (Kohno et al. 2001; Graci\'a-Carpio et al. 2006; Meijerink et al. 2007). In higher metallicity environments, the L(HCN)/L(HCO$^{+}$) ratio varies from 0.5\,-$\geq$4 with a median value of $\sim 1$ (e.g. Graci\'a-Carpio et al. 2006; Baan et al. 2008; Krips et al. 2008). 

The fact that the L(HCN)/L(HCO$^{+}$) ratio is $\sim$1 in high metallicity starbursts likely results from both transitions being optically thick and thermalized. Since HCO$^{+}$ has a critical density almost an order of magnitude lower than HCN (Graci\'a-Carpio et al. 2006; Krips et al. 2008), if the density drops below $\sim$10$^{4-5}$cm$^{-3}$ the L(HCN)/L(HCO$^{+}$) ratio can drop below unity even if optically thick. The effect will be more pronounced if the opacity of HCN and HCO$^{+}$ are lower in the LMC than in these high metallicity starbursts. Thus the small L(HCN)/L(HCO$^{+}$) may suggest that both gas densities and opacities of HCN and HCO$^{+}$ are significantly lower in these LMC GMCs compared to higher metallicity systems. This is reasonable given the lower column densities and lower metallicities of the clumps. Furthermore, there is evidence that the N/O elemental abundance is $\sim$2 - 3 times lower in the LMC than in solar metallicity Galactic environments (e.g. Hunter et al. 2009). If HCN and HCO$^{+}$ abundances reflect the N and O abundances, respectively, this may further explain a portion of the low HCN found in the LMC.

Compared to the L(HCN)/L(HCO$^{+}$) ratio in higher metallicity starburst environments, the ratio of the entire 30Dor-10 GMC is much lower ($\sim$ 0.2-0.3). The $\sim$4 times lower value in the 30Dor-10 (as well as the other LMC GMCs), suggests that the dense gas physical or chemical conditions in the LMC differ from those in massive starburst galaxies. This could be due to the lower density or N/O abundance ratio in the LMC.

Interestingly, there is also no significant difference between the L(HCN)/L(HCO$^{+}$) ratio in 30Dor-10 and the other sampled LMC GMCs (Fig 3d).  Evidently the L(HCN)/L(HCO$^{+}$) ratio is not strongly sensitive to radiation field conditions. This clearly argues against explanations that solely depend on the strength of the radiation field environment.

\section{Conclusion and Summary}

ATCA observations of high dense gas tracers HCO$^{+}$(1-0) and HCN (1-0) in 30Dor-10 were compared to four other star forming GMCs (N159, N113, N105A and N44) in the LMC. Along with the comparison of ALMA CO and MAGMA CO data. Our results are discussed below. 

\begin{itemize}
\item Applying CPROPS we detect 13 HCO$^{+}$ (1-0) bright clumps four of which are detected in HCN(1-0). The clumps exhibit similar masses and sizes but broader linewidths, compared to clumps in the other GMCs in the LMC.  For 30Dor-10 the ratio of $\sum$M$_{\rm vir}$/$\sum$M$_{\rm LTE}$ $\sim$ 6 which is three times the value found for the other four GMCs. 
\end{itemize}

\begin{itemize}
\item All of the results discussed suggest that substructure in 30Dor-10 is not significantly different in mass, dense gas fraction or flux from the other LMC GMCs despite the extreme UV radiation field that surrounds 30Dor-10. It appears that PDRs may not have a dramatic effect on the molecular cloud even with the presence of the super-star cluster, R136. 
\end{itemize}

\begin{itemize}
\item We detect a linear structure that is a potential gravitationally unstable filament and likely collapsing to form stars. The morphology of the filament is in agreement with predictions by the varicose fluid instability where the clumps have roughly uniform spacings. Morphology and dynamics suggest that the filament is lying approximately in the plane of the sky. Toward 30Dor-10, H$_{2}$O masers and YSOs tends to correlate with the positions of HCO$^{+}$ clumps along the filament.   Additional YSOs are observed to extend back toward R136 along the same line. The presence of YSOs and H$_{2}$O masers demonstrates that HCO$^{+}$ cores are currently forming stars. This suggests that the 30Dor-10 clumps may be slightly younger than those near R136 and we are seeing the dispersal of molecular gas in that part of the filament. 
\end{itemize}

\section{Acknowledgments}
Part of this research was conducted at the Jet Propulsion Laboratory, California Institute of Technology under contract with the National Aeronautics and Space Administration. Part of this project was funded by %
NSF Grant AST-1009620 to DSM. This research has made use of the SIMBAD database,
operated at CDS, Strasbourg, France. The National Radio Astronomy Observatory is a facility of the National Science Foundation operated under cooperative agreement by Associated Universities, Inc.

\newpage

\bibliographystyle{apj}

\begin{thebibliography}{23}
\expandafter\ifx\csname natexlab\endcsname\relax\def\natexlab#1{#1}\fi


\bibitem[Andr{\'e} et 
al.(2010)]{2010A&A...518L.102A} Andr{\'e}, P., Men'shchikov, A., Bontemps, S., et al.\ 2010, \aap, 518, L102

\bibitem[Baan et 
al.(2008)]{2008A&A...477..747B} Baan, W.~A., Henkel, C., Loenen, A.~F., Baudry, A., \& Wiklind, T.\ 2008, \aap, 477, 747


\bibitem[Barnes et al.(2011)]{2011ApJS..196...12B} Barnes, P.~J., Yonekura, 
Y., Fukui, Y., et al.\ 2011, \apjs, 196, 12 

\bibitem[Bolatto et 
al.(2013)]{2013ARA&A..51..207B} Bolatto, A.~D., Wolfire, M., \& Leroy, A.~K.\ 2013, \araa, 51, 207

\bibitem[Beltr{\'a}n et 
al.(2006)]{2006A&A...447..221B} Beltr{\'a}n, M.~T., Brand, J., Cesaroni, R., et al.\ 2006, \aap, 447, 221 


\bibitem[Chin et 
al.(1997)]{1997A&A...317..548C} Chin, Y.-N., Henkel, C., Whiteoak, J.~B., et al.\ 1997, \aap, 317, 548


\bibitem[Cohen et al.(1988)]{1988ApJ...331L..95C} Cohen, R.~S., Dame, 
T.~M., Garay, G., et al.\ 1988, \apjl, 331, L95

\bibitem[Draine(1978)]{1978ApJS...36..595D} Draine, B.~T.\ 1978, \apjs, 36, 
595

\bibitem[Fa{\'u}ndez et 
al.(2004)]{2004A&A...426...97F} Fa{\'u}ndez, S., Bronfman, L., Garay, G., et al.\ 2004, \aap, 426, 97

\bibitem[Fukui et al.(1999)]{1999PASJ...51..745F} Fukui, Y., Mizuno, N.,
Yamaguchi, R., et al.\ 1999, \pasj, 51, 745

\bibitem[Graci{\'a}-Carpio et al.(2006)]{2006ApJ...640L.135G} 
Graci{\'a}-Carpio, J., Garc{\'{\i}}a-Burillo, S., Planesas, P., 
\& Colina, L.\ 2006, \apjl, 640, L135 

\bibitem[Gruendl 
\& Chu(2009)]{2009ApJS..184..172G} Gruendl, R.~A., \& Chu, Y.-H.\ 2009, \apjs, 184, 172 

\bibitem[Heikkil{\"a} et 
al.(1999)]{1999A&A...344..817H} Heikkil{\"a}, A., Johansson, L.~E.~B., \& Olofsson, H.\ 1999, \aap, 344, 817

\bibitem[Hughes et al.(2010)]{2010MNRAS.406.2065H} Hughes, A., Wong, T., 
Ott, J., et al.\ 2010, \mnras, 406, 2065

\bibitem[Hunter et 
al.(2009)]{2009A&A...496..841H} Hunter, I., Brott, I., Langer, N., et al.\ 2009, \aap, 496, 841

\bibitem[Indebetouw et al.(2013)]{2013ApJ...774...73I} Indebetouw, R., 
Brogan, C., Chen, C.-H.~R., et al.\ 2013, \apj, 774, 73 

\bibitem[Inutsuka 
\& Miyama(1997)]{1997ApJ...480..681I} Inutsuka, S.-I., \& Miyama, S.~M.\ 1997, \apj, 480, 681

\bibitem[Jackson et al.(2010)]{2010ApJ...719L.185J} Jackson, J.~M., Finn, 
S.~C., Chambers, E.~T., Rathborne, J.~M., 
\& Simon, R.\ 2010, \apjl, 719, L185 

\bibitem[Johansson et 
al.(1994)]{1994A&A...291...89J} Johansson, L.~E.~B., Olofsson, H., Hjalmarson, A., Gredel, R., \& Black, J.~H.\ 1994, \aap, 291, 89

\bibitem[Johansson et 
al.(1998)]{1998A&A...331..857J} Johansson, L.~E.~B., Greve, A., Booth, R.~S., et al.\ 1998, \aap, 331, 857

\bibitem[Kennicutt 
\& Evans(2012)]{2012ARA&A..50..531K} Kennicutt, R.~C., \& Evans, N.~J.\ 2012, \araa, 50, 531

\bibitem[Kohno et al.(2001)]{2001ASPC..249..672K} Kohno, K., Matsushita, 
S., Vila-Vilar{\'o}, B., et al.\ 2001, The Central Kiloparsec of Starbursts 
and AGN: The La Palma Connection, 249, 672 

\bibitem[Krips et al.(2008)]{2008ApJ...677..262K} Krips, M., Neri, R., 
Garc{\'{\i}}a-Burillo, S., et al.\ 2008, \apj, 677, 262

\bibitem[McKee 
\& Ostriker(2007)]{2007ARA&A..45..565M} McKee, C.~F., \& Ostriker, E.~C.\ 2007, \araa, 45, 565

\bibitem[Meier 
\& Turner(2012)]{2012ApJ...755..104M} Meier, D.~S., \& Turner, J.~L.\ 2012, \apj, 755, 104 

\bibitem[Meijerink et 
al.(2007)]{2007A&A...461..793M} Meijerink, R., Spaans, M., \& Israel, F.~P.\ 2007, \aap, 461, 793 

\bibitem[Muller et al.(2010)]{2010ApJ...712.1248M} Muller, E., Ott, J., 
Hughes, A., et al.\ 2010, \apj, 712, 1248

\bibitem[Ostriker(1964)]{1964ApJ...140.1056O} Ostriker, J.\ 1964, \apj, 
140, 1056  

\bibitem[Ott et al.(2008)]{2008PASA...25..129O} Ott, J., Wong, T., Pineda, 
J.~L., et al.\ 2008, PASA, 25, 129

\bibitem[Ott et al.(2010)]{2010ApJ...710..105O} Ott, J., Henkel, C., 
Staveley-Smith, L., \& Wei{\ss}, A.\ 2010, \apj, 710, 105 

\bibitem[Pineda et 
al.(2012)]{2012A&A...544A..84P} Pineda, J.~L., Mizuno, N., R{\"o}llig, M., et al.\ 2012, \aap, 544, A84 

\bibitem[Pineda et al.(2009)]{2009ApJ...703..736P} Pineda, J.~L., Ott, J., 
Klein, U., et al.\ 2009, \apj, 703, 736

\bibitem[Rosolowsky 
\& Leroy(2006)]{2006PASP..118..590R} Rosolowsky, E., \& Leroy, A.\ 2006, PASP, 118, 590

\bibitem[Rubio et 
al.(2009)]{2009A&A...505..177R} Rubio, M., Paron, S., \& Dubner, G.\ 2009, \aap, 505, 177

\bibitem[Sault et al.(1995)]{1995ASPC...77..433S} Sault, R.~J., Teuben, 
P.~J., 
\& Wright, M.~C.~H.\ 1995, Astronomical Data Analysis Software and Systems IV, 77, 433

\bibitem[Seale et al.(2012)]{2012ApJ...751...42S} Seale, J.~P., Looney, 
L.~W., Wong, T., et al.\ 2012, \apj, 751, 42

\bibitem[Strong et 
al.(1988)]{1988A&A...207....1S} Strong, A.~W., Bloemen, J.~B.~G.~M., Dame, T.~M., et al.\ 1988, \aap, 207, 1

\bibitem[Wang et al.(2009)]{2009ApJ...690..580W} Wang, M., Chin, Y.-N., 
Henkel, C., Whiteoak, J.~B., \& Cunningham, M.\ 2009, \apj, 690, 580

\bibitem[Williams et al.(2000)]{2000prpl.conf...97W} Williams, J.~P., 
Blitz, L., \& McKee, C.~F.\ 2000, Protostars and Planets IV, 97 

\bibitem[Wong et al.(2006)]{2006ApJ...649..224W} Wong, T., Whiteoak, J.~B., 
Ott, J., Chin, Y.-n., \& Cunningham, M.~R.\ 2006, \apj, 649, 224

\bibitem[Wong et al.(2011)]{2011ApJS..197...16W} Wong, T., Hughes, A., Ott, 
J., et al.\ 2011, \apjs, 197, 16 

\bibitem[Zinnecker 
\& Yorke(2007)]{2007ARA&A..45..481Z} Zinnecker, H., \& Yorke, H.~W.\ 2007, \araa, 45, 481


\end{thebibliography}

\clearpage

\begin{deluxetable}{ccrrcccccccc}
\scriptsize
\rotate
	\tabletypesize{\scriptsize}
	\tablecaption{Properties of Individual Clumps in 30Dor-10}
	\tablewidth{0pt}
	\tablenum{1}
	\tablehead{\colhead{ATCA}&\colhead{ALMA}&\colhead{R.A.}&\colhead{Dec.}&\colhead{$\Delta$v}&\colhead{R}&\colhead{M$_{\rm vir}$}&\colhead{M$_{\rm LTE}$}&\colhead{L$_{\rm HCO^{+}}$}&\colhead{L$_{\rm HCN}$}&\colhead{L$_{\rm HCN}$/L$_{\rm HCO^{+}}$}&\colhead{$^{a}\rm v_{\rm cent}$}\\
\colhead{Clump ID}&\colhead{Clump ID}&\colhead{[h:m:s]}&\colhead{[$^{o}$:\arcmin:\arcsec]}&\colhead{\rm km s$^{-1}$}&\colhead{[\rm pc]}&\colhead{[$\times 10^{3}$M$_\odot$]}&\colhead{[$\times 10^{3}$M$_\odot$]}&\colhead{[\rm K km s$^{-1}$pc$^{2}$]}&\colhead{[\rm K km s$^{-1}$pc$^{2}$]}&\colhead{}&\colhead{\rm km s$^{-1}$}}

\startdata

1	& - &	05:38:55.09	&	-69:04:16.9	&	2.9(0.6)	&	$<$\,1.0			&	$<$\,2.3				&	$<$\,1.5		&	$<$\,8.4			&	2.1(0.5)		& 0.4(0.5)	&	240.6	\\
2	& - &	05:38:52.17	&	-69:04:06.1	&	2.5(0.6)	&	$<$\,2.0			&	$<$\,2.6				&	$<$\,3.7		&	6.7(4.8)			&	1.4(0.5)		& 0.2(0.2)	&	243.0	\\
3	& 83,85,87-94 &	05:38:50.73	&	-69:04:21.9	&	3.1(0.4)	&$\lesssim$0.7	(0.5)		&	$\lesssim$\,2.7(1.2)			&	$<$\,0.3		&	14.6(7.4)		&	1.7(0.4)		& 0.1(0.06) 	&	246.2	\\
4	& 4-7,9-10&	05:38:45.24	&	-69:05:05.7	&	4.5(0.3)	&	$<$\,0.9			&	$\lesssim$\,5.5	(4.2)		&	$<$\,2.4		&	28.2(4.1)		&	3.0(1.0)		& 0.1(0.04)	&	248.9	\\
5	& 11-14,21-28,30-32,34&	05:38:47.02	&	-69:04:58.9	&	3.4(0.2)	&	$\lesssim$\,0.7 (0.2)	&	$\lesssim$\,3.2	(1.0)		&	0.3(0.1)		&	30.3(3.2)		&	5.4(1.3)		& 0.2(0.05)	&	250.3	\\
6	& 35,50-55,57-58&	05:38:49.03	&	-69:04:43.8	&	6.1(0.1)	&	1.1 (0.2)				&			7.6(2.5)				&	1.5	(0.1)	&	81.4(6.8)		&	16.9(2.4)	& 0.2(0.03)	&	251.0	\\
7	&44-47&	05:38:45.08	&	-69:04:39.0	&	2.7(0.3)	&	$<$\,1.6			&	$<$\,2.1				&	$<$\,4.7		&	21.4(3.4)		&	3.0(0.8)		& 0.1(0.04)	&	251.4	\\
8	&36-43,49&	05:38:46.80	&	-69:04:42.6	&	4.4(0.3)	&	0.9 (0.3)				&			3.3(2.6)				&	0.5(0.2)		&	25.8(3.8)		&	5.5(1.1)		& 0.2(0.05)	&	254.0	\\
9	& - &	05:38:51.34	&	-69:03:15.4	&	1.8(1.6)	&	$<$\,0.7 (0.6)	&	$<$\,1.5				&	$<$\,0.3		&	$<$\,10.1		&	0.1(0.1) 	& 0.02(0.02)	&	247.8	\\
10	& 71-79&	05:38:52.87	&	-69:04:36.8	&	6.9(0.1)	&	$\lesssim$\,0.7	(0.2)	&	$\lesssim$\,12.9(1.9)		&	0.7(0.1)		&	90.3(6.0)		&	26.3(3.7)	& 0.3(0.05)	&	253.9	\\
11	& - &	05:38:47.47	&	-69:04:05.1	&	3.0(0.2)	&	1.4 (0.4)				&			2.4(1.1)				&	0.9	(0.1)	&	33.9(4.6)		&	7.0(1.5)		& 0.2(0.05)	&	248.1	\\
12	& - &	05:38:48.27	&	-69:03:35.5	&	3.1(0.5)	&	1.6 (0.9)				&			2.8(3.6)				&	0.7	(0.2)	&	18.9(6.8)		&	1.6(0.6)		& 0.1(0.04)	&	247.8	\\
13	& - &	05:38:56.40	&	-69:04:17.4	&	4.1(0.2)	&	$<$\,0.8			&	$\lesssim$\,4.5	(2.9)		&	$<$\,2.2		&	33.1(4.0)		&	8.3(1.9)		& 0.2(0.06)	&	250.6	\\

\hline
\enddata

\tablecomments{Table 1:  CPROPS algorithm eclumps was run on the HCO$^{+}$ ATCA data to determine the values in this table for 30Dor-10. There are 13 HCO$^{+}$ clumps found by eclumps with location, size, velocity dispersion, virial mass, and centroid velocity. M$_{\rm LTE}$ and luminosities were calculated following S12.$^{a}$Channel width adopted as uncertainty. Errorbars in parentheses. }

\end{deluxetable}

\clearpage

\begin{deluxetable}{lcccccccccccccc}
	\tabletypesize{\scriptsize}
	\tablecaption{Combined Clump Properties of LMC Star Forming Regions}
	\tablewidth{0pt}
	\tablenum{2}
		\tablehead{\colhead{}&\colhead{Units}&\colhead{30Dor-10}&\colhead{N159}&\colhead{N113}&\colhead{N105A}&\colhead{N44}}

\startdata
R$_{\rm CO}$ & pc & 28.7(0.05) & 14.0(0.19) & 18.3(0.12) & 9.6(0.50) & 12.5(0.33) \\
$\Delta$\rm v$_{\rm CO}$ & km\,s$^{-1}$ & 4.6(0.07) & 2.4(0.17) & 2.6(0.08) & 2.5(0.19) & 2.2(0.16) \\
L$_{\rm CO}$ & 10$^{4}$\rm K\,km\,s$^{-1}$\,pc$^{2}$ & 1.7(0.03) & 1.9(0.25) & 1.7(0.03) & 0.6(0.18) & 1.8(0.23) \\
M$_{\rm CO}$ & 10$^{4}$M$_\odot$ &7.5(0.1) & 21.0(1.1) & 7.6(0.1) & 2.8(0.8) & 7.8(1.0) \\
\# of HCO$^{+}$ clumps &-- & 13 & 27 & 14 & 6 & 8\\
$\sum L_{\rm HCO^{+}}$ & 10$^{2}$\rm K\,km\,s$^{-1}$pc$^{2}$ & 4.0(1.5) & 8.1(2.2) & 7.1(2.6) & 2.6(1.5) & 4.5(2.0) \\
$\sum L_{\rm HCN}$ &10$^{2}$\rm K\,km\,s$^{-1}$pc$^{2}$& 0.8(0.1) & 1.5(0.1) & 2.4(0.2) & 0.8(0.1) & 0.8(0.1) \\
L$_{\rm HCN}$/L$_{\rm HCO^{+}}$ & -- & 0.2(0.1) & 0.2(0.1) & 0.3(0.1) & 0.3(0.2) & 0.2(0.1)\\
$\sum M_{\rm vir}$ &10$^{4}$M$_\odot$& 3.4(0.9) & 5.0(1.7) & 1.6(0.6) & 1.4(0.8) & 4.4(2.2) \\
$\sum M_{\rm LTE}$ &10$^{4}$M$_\odot$& 0.6(0.7) & 2.1(0.3) & 1.1(0.9) & 0.7(0.1) & 1.9(0.2) \\
$\Delta$\rm v$_{\rm rms}$ &\rm km\,s$^{-1}$& 3.9 & 2.6 & 2.0 & 2.2 & 2.1 \\
$\sum L_{\rm HCO^{+}} / L_{\rm CO}$ &--& 0.02 & 0.08 & 0.04 & 0.04 & 0.03 \\
Radiation Field &$\chi/\chi_{0}$& 562 & 156 & 97 & 83 & 76 \\

\enddata

\tablecomments{Table 2: GMC properties of five star forming regions are shown here. The CPROPS algorithm was used on the MAGMA survey for the CO luminosities, linewidths, and radii of each cloud. The radiation field is determined by the flux from HIRAS (IRAS 60$\mu$m and 100$\mu$m high resolution data) where the far infrared intensity and FUV field strength is calculated using Pineda et al. (2009) equations (4) and (5). }

\end{deluxetable}

\clearpage

  \begin{figure}[!ht]
    \begin{center}
     \begin{tabular}{c}
      \begin{minipage}[t]{1.0 \textwidth}
	\vskip 0.0cm
	\hskip 0.0cm
	\centerline{\epsfxsize=1.0\textwidth \epsfbox{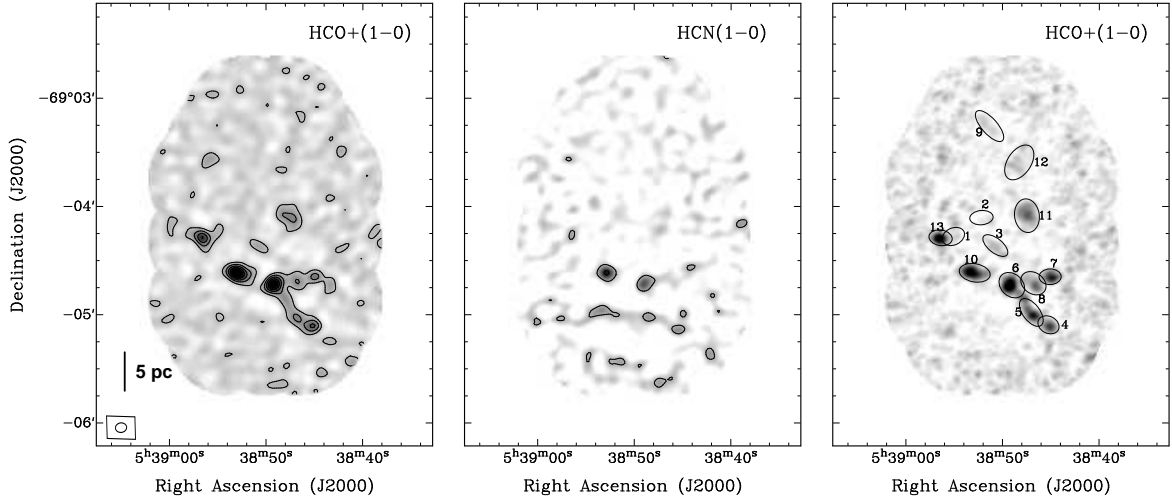}}
  \caption{\textit{Left:} HCO$^{+}$ integrated intensity map of the full ATCA mosaic field of view. Contours in black are set to the lowest value of 0.3 Jy/beam km s$^{-1}$ with 0.6 Jy/beam km s$^{-1}$ increments. \textit{Middle:} HCN integrated intensity map of the full ATCA mosaic field of view. Contours in black are set to the same levels as for the HCO$^{+}$ map.  \textit{Right:} Using the CPROPS algorithm, 13 clumps were detected and their masses, sizes, and linewidths can be found in Table 1. The ellipses found from CPROPS are shown overlaid on the HCO$^{+}$ CPROPS peak temperature image indicating the location of the clumps. The detection of clumps 1, 2, 9, and 12 is tentative.
         \label{fig3}}
       \end{minipage}
     \end{tabular}
    \end{center}
   \end{figure}

   \clearpage
   
   \begin{figure}[!ht]
    \begin{center}
     \begin{tabular}{c}
      \begin{minipage}[t]{1.0\textwidth}
	\vskip -4.0cm
	\hskip 2.0cm
	\centerline{\epsfxsize=2.0\textwidth \epsfbox{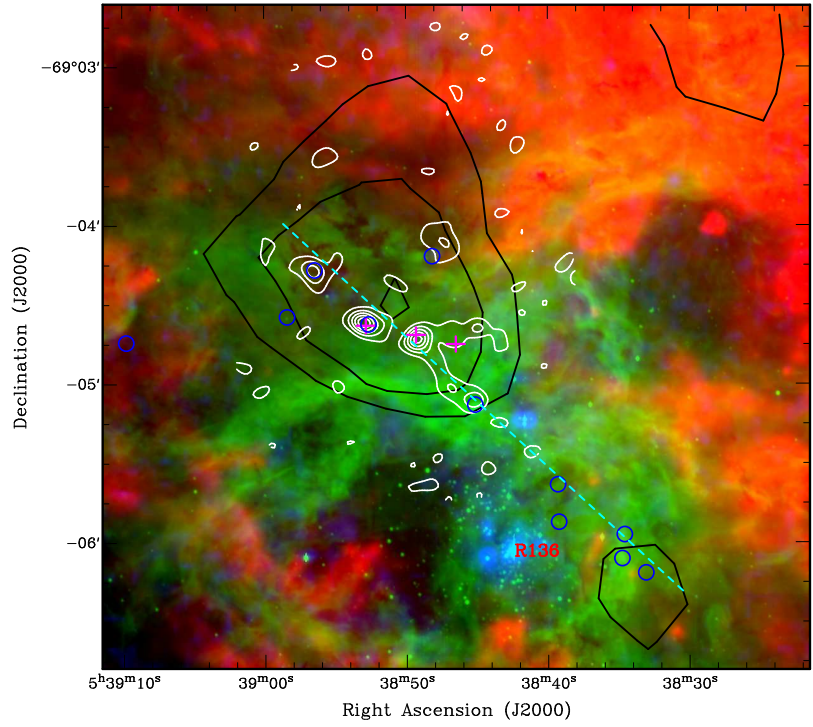}}
	\vskip -2.0cm
  \caption{Image of the 30Doradus region with Spitzer 8$\micron$ in red, MCELS H$\alpha$ in green, and X-rays from Chandra in blue with MAGMA CO(1-0) contours overlaid in grey. Blue circles indicate the location of YSOs. Magenta marks the locations of the H$_{2}$O masers. YSO and maser locations lie along a possible filamentary structure represented by the dashed line. White contours represent the ATCA HCO$^{+}$ integrated intensity map also shown in Figure 1. 
	         \label{fig2}}
       \end{minipage}
     \end{tabular}
    \end{center}
   \end{figure}

\clearpage

\begin{figure}[!ht]
    \begin{center}
     \begin{tabular}{c}
      \begin{minipage}[t]{1.0 \textwidth}
	\vskip  0cm
	\hskip 0.0cm
	\centerline{\epsfxsize=1.0\textwidth \epsfbox{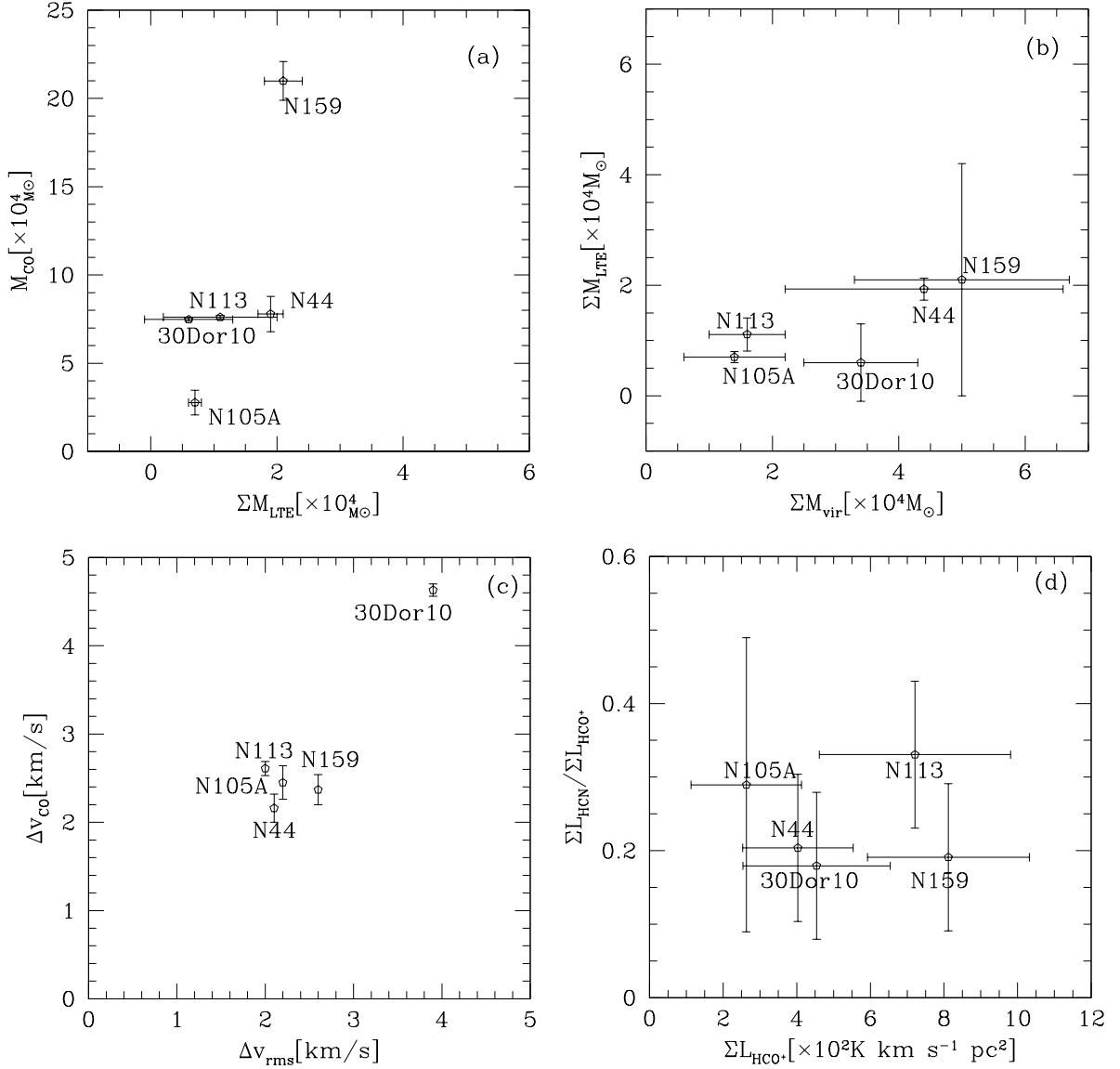}}
    \caption{a) Each GMCs' H$_{2}$ mass traced by HCO$^{+}$ obtained from CPROPS versus the MAGMA CO mass is plotted. b)The virial mass versus the sum of each GMC's H$_{2}$ mass traced by HCO$^{+}$ obtained from CPROPS is plotted here for each GMC and labeled accordingly. c)The root mean square (RMS) of the centroid velocity of the clumps in each GMC versus the CO velocity of the GMCs is plotted. d)Plotted here is sum of all of the clumps in each GMC HCO$^{+}$ luminosities versus the line ratio HCN/HCO$^{+}$. 
       \label{fig3}}
       \end{minipage}
     \end{tabular}
    \end{center}
   \end{figure}

 \clearpage  
 
  \begin{figure}[!ht]
    \begin{center}
     \begin{tabular}{c}
      \begin{minipage}[t]{1.0 \textwidth}
	\vskip  0cm
	\hskip -1.0cm
	\centerline{\epsfxsize=0.7\textwidth \epsfbox{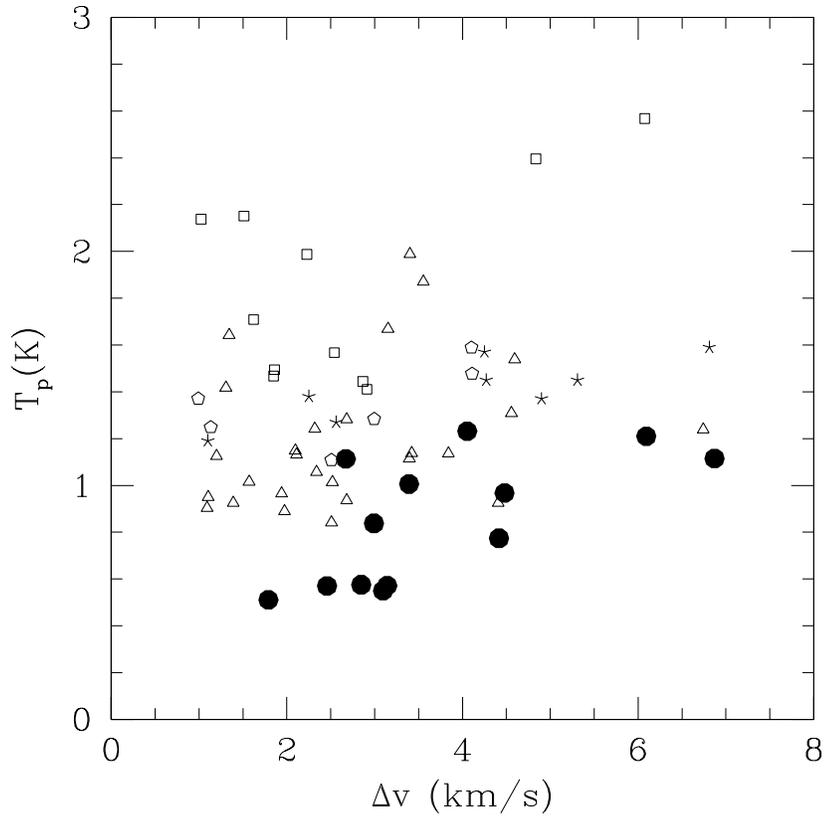}}
 \caption{The linewidth for each clump in the corresponding GMC versus the peak temperature is plotted. Each GMC is represented by a different symbol in this plot. 30Dor-10 = bold filled circles, N159 = triangles, N113 = circles, N105 = squares, N44 = stars.   
       \label{fig3}}
       \end{minipage}
     \end{tabular}
    \end{center}
   \end{figure}
       
\clearpage

 \begin{figure}[!ht]
    \begin{center}
     \begin{tabular}{c}
      \begin{minipage}[t]{1.0 \textwidth}
	\vskip  0cm
	\hskip 0.0cm
	\centerline{\epsfxsize=1.0\textwidth \epsfbox{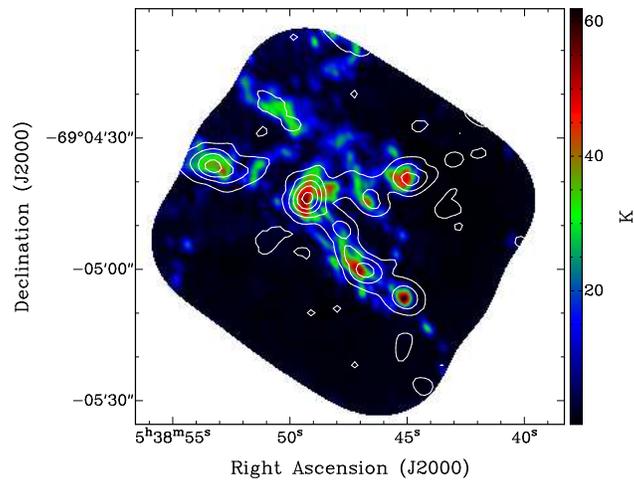}}
\parbox[]{6.5in}{
  \caption{Peak temperature map of ALMA $^{12}$CO(2-1) (Indebetouw et al. 2013) with ATCA HCO$^{+}$ peak temperature contours overlaid. Max contour set at 2.2\,K with contour increments at 20\%, 40\%, 60\% and 80\% of the max. 
       \label{fig3}}}
       \end{minipage}
     \end{tabular}
    \end{center}
   \end{figure}

\clearpage

 \begin{figure}[!hb]
    \begin{center}
     \begin{tabular}{c}
      \begin{minipage}[t]{1.0 \textwidth}
	\vskip  0cm
	\hskip .0cm
	\centerline{\epsfxsize=0.5\textwidth \epsfbox{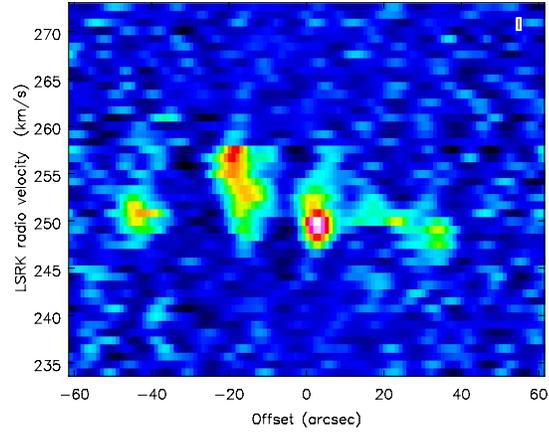}}
  \caption{The position-velocity structure of the filament is determined by taking a slice along the spatial axis of the HCO$^{+}$ cube. The slice used on the HCO$^{+}$ cube had a position angle of 53.44$^{\rm o}$ and a length of 1.841$^{\prime}$ with an averaging width of 19 pixels centered at a RA and DEC of 05:38:49.7, -69:04:40.9. 
       \label{fig3}}
       \end{minipage}
     \end{tabular}
    \end{center}
   \end{figure}

\end{document}